%% file: robust-EM-Clustering-gMixReg.tex
\documentclass[conference]{IEEEtran}

\usepackage{graphicx, epsfig,times,amsmath,amsfonts,amssymb,array,enumerate}
\usepackage{algorithm, algorithmic} 
\usepackage[subnum]{cases}
\usepackage[T1]{fontenc}

\input{mymathsym.tex}

\graphicspath{{Figures/}}

\IEEEoverridecommandlockouts    

\begin{document}

\title{\ \\ \LARGE\bf  Robust EM algorithm for model-based curve clustering 
 \thanks{
Faicel Chamroukhi is with the Information Sciences and Systems Laboratory (LSIS), UMR CNRS  7296 and the University of the South Toulon-Var (USTV). Contact: faicel.chamroukhi@univ-tln.fr}}

\author{Faicel Chamroukhi} 
\maketitle

\begin{abstract}
Model-based clustering approaches concern the paradigm of exploratory data analysis relying on the finite mixture model to automatically find a latent structure governing observed data. They are one of the most popular and successful approaches in cluster analysis. The mixture density estimation is generally performed by maximizing the observed-data log-likelihood by using the expectation-maximization (EM) algorithm. However, it is well-known that the EM algorithm initialization is crucial. In addition, the standard EM algorithm requires the number of clusters to be known a priori. Some solutions have been provided in \cite{Robust_EM_GMM, FigueiredoUnsupervisedlearningMixtures} for model-based clustering with Gaussian mixture models for multivariate data. In this paper we focus on model-based curve clustering approaches, when the data are curves rather than vectorial data, based on regression mixtures. We propose a new robust EM algorithm for clustering curves. We extend the model-based clustering approach presented in \cite{Robust_EM_GMM} for Gaussian mixture models, to the case of curve clustering by regression mixtures, including  polynomial regression mixtures as well as spline or B-spline regressions mixtures. Our approach both handles the problem of initialization and the one of choosing the optimal number of clusters as the EM learning proceeds, rather than in a two-fold scheme. This is achieved by optimizing a penalized log-likelihood criterion. A simulation study confirms the potential benefit of the proposed algorithm in terms of robustness regarding initialization and funding the actual number of clusters.
\end{abstract}

\section{Introduction}

Clustering approaches concern the paradigm of exploratory data analysis in which we aim at automatically finding a latent structure within an observed dataset through which the data can be summarized in a few number of clusters. 
From a statistical learning prospective, cluster analysis involves unsupervised learning techniques, in the sense that the class labels of the data are unknown (missing, hidden or difficult to obtain), to learn an underlying structure of the latent data governing the observed data. This is generally performed through learning the parameters of a latent variable model. 
Clustering techniques are therefore suitable for many application domains   where labeled data is difficult to obtain, or to explore and characterize a dataset before running a supervised learning algorithm, etc.
The aim of clustering in general is to find a partition of the data by dividing them into clusters (groups) such that the data within the same group tend to be more similar, in the sense of a chosen dissimilarity measure, to one another as compared to the data belonging to different groups. 
The definition of (dis)similarity and the method in which the data are clustered differs based on the clustering algorithm being applied. 
One can distinguish four categories of clustering algorithms: hierarchical clustering, distance-based (or prototype-based) clustering, topographic clustering and model-based clustering.
Hierarchical clustering \cite{hastieTibshiraniFreidman_book_2009} aims at building a hierarchy of clusters and includes ascending (or agglomerative) hierarchical clustering and descending (or splitting) hierarchical clustering. In the former, each data example starts in its own cluster, and pairs of clusters are successively merged as one moves up the hierarchy. The pair of clusters to be merged is chosen as the pair
of closest clusters in the sense of a chosen distance criterion (e.g. Ward distance, etc). 
The latter approach operates in the other sense. 
Among one of the most known prototype-based clustering algorithms, one can cite the $K$-means algorithm \cite{MacQueen_Kmeans, HartiganAndWong:Kmeans}. $K$-means is a straightforward and widely used algorithm 
in cluster analysis. It is an iterative clustering algorithm that partitions a given dataset into a predefined number of clusters $K$ by minimizing the within-cluster variance criterion (intra-class inertia). Several variants of $K$-means, including fuzzy $K$-means \cite{Bezdek_fuzzy_Kmeans}, trimmed $K$-means \cite{TrimmedKmeans, PropertiesKmeansAndTrimmedKmeans}, etc, have been proposed. 
One of the most popular topographic clustering approaches is the Self Organizing Map \cite{kohonen_SOM}. The SOM is an unsupervised neural-based approach for data clustering and visualization. It generalizes the competitive learning \cite{kong_competitive_learning_91} by allowing also the neighbours of the winner to be updated. 
This is performed by minimizing a cost function (distance criterion) taking into account the topological aspect of the data through a neighborhood kernel (e.g. Gaussian).
%
%
These clustering approaches can be seen as deterministic as they do not define a density model on the data.
When the clustering approaches rely on density modeling, clustering is generally performed  based on the finite mixture model \cite{mclachlanFiniteMixtureModels}. This approach is known as the model-based clustering \cite{banfield_and_raftery_93, mclachlan_basford88, Fraley2002_model-basedclustering}. Mixture model-based approaches are indeed one of the most popular and successful unsupervised learning approaches in cluster analysis.
In the finite mixture approach for cluster analysis, the data probability density function is assumed to be a mixture density, each component density being associated with a cluster. The problem of clustering therefore becomes the one of estimating the parameters of the assumed mixture model (e.g, estimating the mean vector and the covariance matrix for each component density in the case of Gaussian mixture models). 
The mixture density estimation is generally performed by maximizing the observed data log-likelihood. This can be achieved by the well-known expectation-maximization (EM) algorithm \cite{dlr, mclachlanEM}. 
The EM algorithm is in the core of these model-based clustering approaches thanks to its good desirable properties of stability and reliable convergence.
The EM algorithm is indeed a broadly applicable approach to the iterative computation of maximum likelihood estimates in the framework of latent data models and in particular in finite mixture models. It has a number of advantages, including its numerical stability, simplicity of implementation and reliable convergence. For more account on EM, the reader is referred to \cite{mclachlanEM}. 
%
%
Furthermore,  model-based clustering indeed provides a more general well-established probabilistic framework for cluster analysis compared to deterministic clustering algorithms such as $K$-means algorithms, SOM algorithms, etc. 
For example it has been shown that from a probabilistic point of view, $K$-means is equivalent to a particular case of the Classification EM (CEM) algorithm \cite{celeuxetgovaert92_CEM} for a mixture of $K$ Gaussian densities with the same mixing proportions and identical isotropic covariance matrices. 
The generative version of the Self Organizing Map, that is the Generative Topographic Mapping \cite{bishop_etal_GTM}, allows to overcome the SOM limitations through the probabilistic formulation of a latent variable where both convergence and topographic ordering are guaranteed thanks to the good properties of the EM algorithm.
For all these approaches, the choice of the number of classes can be performed  afterwards the learning process. For example, in model-based clustering, one can use some information criteria for model selection, such as BIC \cite{BIC}, etc, in order to estimate the optimal number of clusters. 

In this paper we focus on model-based clustering approaches, in particular model-based curve clustering based on regressions mixtures. 
Indeed, the main model-based clustering approaches are concerned with vectorial data where the observations are vectors of reduced dimension and the clustering is performed by Gaussian mixture models and the EM algorithm \cite{mclachlanFiniteMixtureModels, mclachlanEM, Robust_EM_GMM}. Each cluster is represented by its mean vector and its covariance matrix.
In many areas of application, 
the data are curves or functions rather than vectors. The analysis approaches are therefore linked to  Functional Data Analysis (FDA) \cite{ramsayandsilvermanFDA2005}
Statistical approaches for FDA \cite{ramsayandsilvermanFDA2005} concern the paradigm of data analysis for which the individuals are entire functions or curves rather than vectors of reduced dimensions.
The goals of FDA, as in classical data analysis, include data representation, regression, classification, clustering, etc. 
From a statistical learning prospective of FDA, the curve clustering can be achieved by learning adapted statistical models, in particular latent data models, in an unsupervised context.  
When the observations are curves or time series, the clustering can therefore be performed model-based curve clustering approaches, namely the regression mixture model 
\cite{GaffneyThesis, chamroukhi_PhD_2010}.
including polynomial regression mixtures, splines and B-splines regression mixtures \cite{GaffneyThesis, chamroukhi_PhD_2010}, or also generative polynomial piecewise regression \cite{chamroukhi_adac_2011, chamroukhi_PhD_2010, chamroukhi_et_al_NN2009}, 
The parameter estimation is still performed by maximizing the observed-data log-likelihood through the EM algorithm.
However, it is well-known that, for Gaussian mixtures as well as for regression mixtures, the initialization of the EM algorithm is a crucial point since it maximizes locally the log-likelihood. 
Therefore, if the initial value is inappropriately selected, the EM algorithm may lead to an unsatisfactory estimation. 
%
In addition, the standard EM algorithm requires the number of clusters to be known which is not always the case. Choosing the optimal number of clusters can be performed afterwards using some information criteria such as the Bayesian information criterion (BIC) \cite{BIC}, 
etc.
%
%
%

In this paper we focus on model-based curve clustering using regressions mixtures. We consider the problem of curve clustering where the observations are temporal curves rathen than vectors as in multivariate Gaussian mixture analysis \cite{Robust_EM_GMM}. We extend the model-based clustering approach presented in \cite{Robust_EM_GMM}, in which a robust EM is developed for Gaussian mixture models, to the case of regression mixtures, including spline regression mixtures or B-spline regression mixtures for curve clustering.
%
%
%
More specifically, we propose a new robust EM clustering algorithm for regressions mixtures, which can be used for polynomial regression mixture as well as for spline or B-spline regressions mixtures. 
Our approach handles both the problem of initialization and the one of choosing the optimal number of clusters as the EM learning proceeds, rather that in a two-fold scheme.
The approach allows therefore for fitting regression mixture models without running an external algorithm for initializing the EM algorithm and without specifying the number of the clusters in the dataset a prior or performing a model selection procedure once the learning has been performed.
We mainly focus on generative approaches which may help us to understand the process generating the curves. The generative approaches for functional data are essentially based on regression analysis, including polynomial 
regression, splines and B-splines \cite{GaffneyThesis, chamroukhi_PhD_2010}. 
%
%
%
 
This paper is organized as follows. 
In the next sections we give a brief background on model-based clustering of multivariate data using Gaussian mixtures and model-based curve clustering using regression mixtures.
Then, in section \ref{sec: proposed robust EM-Mix-Reg} we present the proposed robust EM algorithm
which maximizes a penalized log-likelihood criterion for regression mixture model-based  curve clustering. 
and derive the corresponding parameter updating formulas. 
 

\section{Background on model-based clustering with Gaussian mixtures}
\label{sec: related work MBC}
In this section, we give a brief background on model-based clustering of multivariate data using Gaussian mixtures.
Model-based clustering \cite{banfield_and_raftery_93,mclachlan_basford88,Fraley2002_model-basedclustering}, generally used for multidimensional data, is based on a finite mixture model formulation \cite{mclachlanFiniteMixtureModels}. 

Let us denote by $(\bx_1,\ldots,\bx_n)$ an observed i.i.d  dataset, each observation  $\bx_i$ is represented as multidimensional vector in $\R^d$. 
We let also $\bz = (z_1,\ldots,z_n)$ denotes the corresponding unobserved
(missing) labels where the class label $z_i$ takes its values in the finite set $\{1,\ldots,K\}$, $K$ being the number of clusters.
In the finite mixture  model, the data probability density function is assumed to be a Gaussian mixture density defined as
\begin{equation}
f(\bx_i;\bsPsi) = \sum_{k=1}^K \pi_{k} \N \big(\bx_i;\bsmu_k,\bsSigma_k \big),
\label{eq: GMM}
\end{equation}each component Gaussian density being associated with a cluster. $\N(.;\bsmu,\bsSigma)$ denotes the multivariate Gaussian density with mean vector $\bsmu$ and covariance matrix $\bsSigma$. 
 In this mixture density, the $\pi_k$'s are the non-negative mixing proportions that sum to 1 and $\bsmu_k$ and $\bsSigma_k$ are respectively  the mean vector and the covariance matrix for each mixture component density.
 The problem of clustering therefore becomes the one of estimating the parameters of the Gaussian mixture model $\bsPsi = (\pi_1,\ldots,\pi_K,\bsPsi_1,\ldots,\bsPsi_K)$  where $\bsPsi_k = (\bsmu_k,\bsSigma_k)$. 
This can be performed by maximizing the following observed-data log-likelihood of $\bsPsi$  by using the EM algorithm \cite{mclachlanEM, dlr}: 
\begin{equation}
\cL(\bsPsi) = \sum_{i=1}^{n}\log\sum_{k=1}^K \pi_{k} \N \big(\bx_i;\bsmu_k,\bsSigma_k \big).
\label{eq: loglik normal mixture}
\end{equation}
However, the EM algorithm for Gaussian mixture models is quite sensitive to initial values. The Gaussian mixture model-based clustering technique might yield poor clusters if the mixture parameters are not initialized properly.
The EM initialization can be performed from a randomly chosen partition of the data or by computing a  partition from another clustering algorithm such as $K$-means, Classification EM \cite{celeux_et_diebolt_SEM_85}, Stochastic EM \cite{celeuxetgovaert92_CEM}, etc
or by initializing EM with a few number of steps of EM itself.
%
%
Several works have been proposed in the literature in order to overcome this drawback and  making the EM algorithm for Gaussian mixtures robust with regard initialization \cite{biernacki_etal_startingEM_CSDA03, Reddy:2008, Robust_EM_GMM}. 
Further details about choosing starting values for the EM algorithm for Gaussian mixtures can be found for example in \cite{biernacki_etal_startingEM_CSDA03}. 
On the other hand, the number of mixture components (clusters) needs to be known a priori.
Some authors have considered this issue in order to estimate the unknown number of mixture components, for example as in \cite{Richardson_BayesianMixtures, Robust_EM_GMM}.
In general, theses two issues have been considered each separately. Among the approaches considering both the problem of robustness with regard to initial values and automatically estimating the number of mixture components in the same algorithm, one can cite the EM algorithm proposed in \cite{FigueiredoUnsupervisedlearningMixtures}.
In \cite{FigueiredoUnsupervisedlearningMixtures}, the authors proposed an EM algorithm that is capable of selecting the number of components and it attempts to be not sensitive with regard to initial values.
In deed, the algorithm developed in \cite{FigueiredoUnsupervisedlearningMixtures} optimizes a particular criterion called the minimum message length (MML), which is a penalized negative log-likelihood rather than the observed data log-likelihood. The penalization term allows to control the model complexity. Indeed, the EM algorithm in \cite{FigueiredoUnsupervisedlearningMixtures} starts by fitting a mixture model with large number of clusters and discards illegitimate clusters as the learning proceeds.
The degree of legitimate of each cluster is measured through the penalization term which includes  the mixing proportions to know if the cluster is small or not to be discarded and therefore to reduce the number of clusters.
%
%

%
%
More recently, in \cite{Robust_EM_GMM}, the authors developed a robust EM algorithm for model-based clustering of multivariate data using Gaussian mixture models that simultaneously addresses the problem of initialization and estimating the number of mixture components. This algorithm overcome some initialization drawback of the EM algorithm proposed in \cite{FigueiredoUnsupervisedlearningMixtures}, which is still having an initialization problem. As shwon in  \cite{Robust_EM_GMM}, this problem can become more serious especially for a dataset with a large number of clusters. 
For more details on robust EM clustering for Gaussian mixture models for multivariate data, the reader is referred to \cite{Robust_EM_GMM}.

However, these presented model-based clustering approaches, namely \cite{Robust_EM_GMM} \cite{FigueiredoUnsupervisedlearningMixtures}, are concerned with vectorial data where the observations are assumed to be vectors of reduced dimension.
When the data are rather curves or functions, one can rely on regression mixtures  \cite{GaffneyThesis, chamroukhi_PhD_2010} or generative hidden process regression \cite{chamroukhi_et_al_neurocomputing2010}\cite{chamroukhi_PhD_2010} which are more adapted than standard Gaussian mixture modeling.

In the next section we describe the model-based curve clustering approach based on regression mixtures using the standard EM algorithm and then we derived our EM algorithm in the following section. 
  


\section{Background on model-based curve clustering with regression mixtures}
\label{sec: background on MBCC}


\subsection{Model-based curve clustering}

Mixture-model-based clustering approaches have also been introduced to generalize the standard multivariate mixture model to the analysis of curves where the individuals are presented as curves rather than a vector of a reduced dimension.  
In this case,  one can distinguish the  regression mixture approaches, including polynomial regression and spline regression, or random effects polynomial regression \cite{GaffneyThesis, garetjamesJASA2003}. 
Notice that the mixture of regression models are similar to the well-known mixture of experts (ME) model \cite{jacobsME}. Although similar, mixture of experts differ from curve clustering
models in many respects (for instance see \cite{GaffneyThesis}). One of the main differences is that the ME model consists in a conditional mixture while the mixture of regressions is an unconditional one. Indeed, the mixing proportions are constant for the mixture of regressions, while in the mixture of experts, they mixing proportions (known as the gating functions of the network) are modeled as a function of the inputs, generally as a logistic or a softmax function.  
All these approaches use the mixture (estimation) approach with the EM algorithm to estimate the model parameters. 
One can also consider generative hidden process regression mixtures \cite{chamroukhi_et_al_neurocomputing2010}\cite{chamroukhi_PhD_2010} \cite{chamroukhi_adac_2011} which performs curve segmentation.

  
\subsection{Regression mixtures for model-based curve clustering} 
\label{ssec: polynomial and spline regression mixture}
 

In this section we describe related model-based curve clustering approaches based on regression mixture models 
\cite{GaffneyThesis, chamroukhi_PhD_2010}. 
%

%
%
The aim of EM clustering in the case of regression mixtures is to cluster $n$ iid curves $((\bx_1,\by_1),\ldots,(\bx_n,\by_n))$ into $K$ clusters. 
We assume that each curve consists of $m$ observations $\by_i=(y_{i1},\ldots,y_{im})$ regularly observed at the inputs $\bx_i = (x_{i1},\ldots,x_{im})$ for all $i=1,\ldots,n$ (e.g., $\bx$ may represent the sampling time in a temporal context).
Let  $\bz = (z_1,\ldots,z_n)$ be the unknown cluster labels associated with the set of curves (time series) $(\by_1,\ldots,\by_n)$, with $z_i \in \{1,\ldots,K\}$, $K$ being the number of clusters.
The regression mixtures approaches assume that each curve is drawn from one of $K$ clusters of curves whose proportions (prior probabilities) are $(\pi_1,\ldots,\pi_K)$. Each cluster of curves is supposed to be a set of homogeneous curves modeled by for example a  polynomial regression model or a spline regression model.  
%
%
The polynomial Gaussian regression mixture model arises when we assume that, the curve $(\bx_i,\by_i)$ has a prior probability $\pi_k$ to be generated from the cluster $z_i = k$  and is generated according  to a noisy polynomial (or spline) function with (polynomial) coefficients $\bsbeta_k$ corrupted by a standard zero-mean Gaussian noise with a covariance-matrix $\sigma_k^2\Identity_m$, that is, for the cluster $k$ we have
\begin{equation}
\by_i = \bX_i \bsbeta_k + \bsepsilon_i
\label{eq: polynomial regression model}
\end{equation}
where $\by_i = (y_{i1},\ldots,y_{im})^T$ is an $m$ by $1$ vector, $\bX_i$ 
is the $m$ by $(p+1)$ regression matrix (Vandermonde matrix) with rows $\bsx_i = (1, x_{ij},x_{ij}^{2} \ldots, x_{ij}^{p})$, $p$ being the polynomial degree, $\bsbeta_k=(\beta_{k0},\ldots,\beta_{kp})^T$ is the  $(p+1)$ by $1$ vector of regression coefficients for class $k$, 
 $\bsepsilon_i \sim \N(\bO,\sigma_k^2\Identity_m)$ is a multivariate standard zero-mean Gaussian variable with a variance $\sigma_k^2\Identity_m$,
 representing the corresponding Gaussian noise and $\Identity_m$ denotes the $m$ by $m$ identity matrix.  
We notice that the regression mixture is adapted for polynomial regression mixture as well as for spline of B-slipne regression mixtures. This only depends on the construction of the regression matrix $\bX_i$.	
The conditional density of curves from cluster $k$ is therefore given by
\begin{equation}
f_k(\by_i|\bx_i, z_i = k;\bsbeta_k,\sigma_k^2) = \N (\by_{i};\bX_i \bsbeta_k,\sigma_k^2\Identity_m), 
\label{eq: polynomial regression density model}
\end{equation}
and finally the conditional mixture density of the $i$th curve can be written as: 
\begin{IEEEeqnarray}{lll}
f(\by_i|\bx_i;\bsPsi) &=& \sum_{k=1}^K p(z_i=k)f_k(\by_i|\bx_i, z_i = k;\bsbeta_k,\sigma_k^2) \IEEEnonumber \\
 &= &\sum_{k=1}^K \pi_k \  \N (\by_{i};\bX_i \bsbeta_k,\sigma_k^2\Identity_m). 
\label{eq: polynomial regression mixture (PRM)}
\end{IEEEeqnarray}where the model parameters are given by 
 $\bsPsi = (\pi_1,\ldots,\pi_K,\bsPsi_1,\ldots,\bsPsi_K)$ where
the $\pi_k$'s are the non-negative mixing proportions that sum to 1
 $\bsPsi_k=(\bsbeta_{k},\sigma_{k}^2)$, $\sigma_{k}^2$ represents the regression parameters and the noise variance for cluster $k$. The unknown parameter vector $\bsPsi$ is generally estimated by maximizing the observed-data log-likelihood of $\bsPsi$.
 Given an i.i.d training set of $n$ curves the  observed-data log-likelihood of $\bsPsi$ is given by:
\begin{equation}
\cL(\bsPsi) = 
\sum_{i=1}^n  \log \sum_{k=1}^K \pi_k \ \N (\by_{i};\bX_i \bsbeta_k,\sigma_k^2\Identity_m).
\label{eq: log-lik PRM}
\end{equation}
and its maximization is performed iteratively via the EM algorithm \cite{GaffneyThesis, dlr, chamroukhi_PhD_2010}. 

  \subsection{Standard EM algorithm for regression mixtures}
  \label{ssec: EM_MixReg}
The maximization of the observed-data the log-likelihood (\ref{eq: log-lik PRM}) by the EM algorithm relies on the complete-data log-likelihood, that is assuming the hidden labels $\bz$ are known:
\begin{IEEEeqnarray}{lcl}
\cL_c(\bsPsi) & = & \sum_{i=1}^{n}\sum_{k=1}^{K} z_{ik} \log \big[\pi_k \N  (\by_{i};\bX_i\bsbeta_{k},\sigma^2_{k}\Identity_m)\big] 
 \label{eq: complete log-lik PRM}
\end{IEEEeqnarray}
where $z_{ik}$ is an indicator binary-valued variable such that $z_{ik}=1$ if $z_i=k$ (i.e., if $\by_i$ is generated by the polynomial regression component $k$) and $z_{ik}=0$ otherwise.

At each iteration,  the EM maximizes the expectation of the complete-data log-likelihood (\ref{eq: complete log-lik PRM}) given the observed data $\cD = ((\bx_1,\by_1),\ldots,(\bx_n,\by_n))$ and a current parameter estimation of $\bsPsi$. 
The EM algorithm for regression mixtures therefore starts with an initial model parameters $\bsPsi^{(0)}$ and alternates between the two following steps until convergence:
\subsubsection{E-step}
\label{ssec: E-step EM-MixReg}

Compute the expected complete-data log-likelihood given the observed curves $\cD = ((\bx_1,\by_1),\ldots,(\bx_n,\by_n))$ and the current value of the parameter $\bsPsi$ denoted by  $\bsPsi^{(q)}$, $q$ being the current iteration number:  
{\begin{IEEEeqnarray}{lll}
Q(\bsPsi,\bsPsi^{(q)})&=& \E\big[\cL_c(\bsPsi)|\cD;\bsPsi^{(q)}\big]
\IEEEnonumber \\
&=& \sum_{i=1}^{n}\sum_{k=1}^{K}\tau_{ik}^{(q)} \log \big[\pi_k  \N  (\by_i;\bX \bsbeta_{k},\sigma^2_{k}\Identity_m)\big]\cdot
\label{eq: Q-function PRM}
\end{IEEEeqnarray}}This simply consists in computing the posterior probability that the $i$th curve  is generated from cluster $k$
\begin{IEEEeqnarray}{lcl}
\tau_{ik}^{(q)}& = & p(z_i=k|\by_{i},\bx_i;\bsPsi^{(q)}) \IEEEnonumber \\
& = & \frac{\pi_k^{(q)} \N\big(\by_i;\bX_i \bsbeta^{(q)}_{k},\sigma^{2(q)}_{k}\Identity_m\big)}{\sum_{h=1}^K \pi_{h}^{(q)} \N(\by_i;\bX_i \bsbeta^{(q)}_{h},\sigma^{2(q)}_{h}\Identity_m)} 
\label{eq: post prob tauik PRM}
\end{IEEEeqnarray}
for each curve and for each of the $K$ clusters.  

\subsubsection{M-step}
\label{ssec: M-step EM-MixReg}This step updates the models parameters by computing the update $\bsPsi^{(q+1)}$ by maximizing the $Q$-function (\ref{eq: Q-function PRM})  with respect to $\bsPsi$.  
The updating formula for the mixing proportion is obtained by maximizing
$\sum_{i=1}^{n}\sum_{k=1}^{K} \tau^{(q)}_{ik} \log \pi_k$ with respect to $(\pi_1,\ldots,\pi_R)$ subject to the constraint $\sum_{k=1}^K \pi_k = 1$ using Lagrange multipliers which gives the following updates \cite{mclachlanEM, dlr, GaffneyThesis, chamroukhi_PhD_2010}
\begin{IEEEeqnarray}{lcl}
\pi_k^{(q+1)} &=& \frac{1}{n}\sum_{i=1}^n \tau^{(q)}_{ik} \quad (k=1,\ldots,K).
\label{eq: EM-MixReg pik update}
\end{IEEEeqnarray}
The maximization of each of the $K$ functions $\sum_{i=1}^{n} \tau_{ik}^{(q)} \log  \N  (\by_i;\bX \bsbeta_{k},\sigma^2_{k}\Identity_m)$ consists in solving a weighted least-squares problem. The solution of this problem is straightforward and these update equations are equivalent to the well-known weighted least-squares solutions (see for example \cite{GaffneyThesis, chamroukhi_PhD_2010}):
\begin{IEEEeqnarray}{lcl}
\bsbeta_k^{(q+1)}  &=& \Big[\sum_{i=1}^{n}\tau^{(q)}_{ik} \bX^T_i\bX_i \Big]^{-1} \sum_{i=1}^{n}\tau^{(q)}_{ik} \bX_i^T \by_i
\label{eq: EM-MixReg beta_k update PRM}
\end{IEEEeqnarray} 
\begin{IEEEeqnarray}{lcl}
\sigma_k^{2(q+1)} &=& \frac{1}{\sum_{i=1}^{n}\tau^{(q)}_{ik}} \sum_{i=1}^{n}\tau^{(q)}_{ik} \parallel \by_i - \bX_i\bsbeta_k\parallel^2
\label{eq: EM-MixReg sigma_k update PRM}
\end{IEEEeqnarray}
Once the model parameters have been estimated, a partition of the data into $K$ clusters is then computed by maximizing the posterior cluster probabilities (\ref{eq: post prob tauik PRM}):
\begin{IEEEeqnarray}{lcl}
\hat{z}_i = \arg \max_{k=1}^K \hat \tau_{ik} 
\end{IEEEeqnarray}
%
 %
However, it can be noticed that, the standard EM algorithm for regression mixture model is sensitive to initialization. In addition, it requires the number of clusters to be supplied by the user. While the number of cluster can be chosen by some model selection criteria, this requires performing additional model selection procedure.
In this paper, we attempt to overcome these limitations in this case of model-based curve clustering by proposing an EM algorithm which is robust with regard initialization and automatically estimate the optimal number of clusters as the learning proceeds. 
%
  
\section{Proposed robust EM algorithm for model-based curve clustering}
\label{sec: proposed robust EM-Mix-Reg}
In this section we present the proposed EM algorithm for model-based curve clustering using regression mixtures.
The present work is in the same spirit of the EM algorithm presented in  \cite{Robust_EM_GMM}
 but by extending the idea to the case of curve clustering rather than multivariate data clustering. Indeed, the data here are assumed to be curves rather than vectors of reduced dimensions.
This leads to fitting a regression mixture model (including splines or B-splines), rather than fitting standard Gaussian mixtures. 
We start by describing the maximized objective function and then we derive the proposed EM algorithm to estimate the regression mixture model parameters.

\subsection{Penalized maximum likelihood estimation}
For estimating the regression mixture model \ref{eq: polynomial regression mixture (PRM)}, we attempt to maximize a penalized log-likelihood function rather than the standard observed-data log-likelihood (\ref{eq: log-lik PRM}). 
%
%
This criterion consists in penalizing the observed-data log-likelihood (\ref{eq: log-lik PRM}) by a term accounting for the model complexity. As the model complexity is governed by in particular the number of clusters and thefore the structure of the hidden variables $z_i$. We chose to use as penalty the entropy of the hidden variable $z_i$ (we recall that $z_i$ is the class label of the $i$th curve).
The penalized log-likelihood criterion is therefore derived as follows.
The discrete-valued variable $z_i \in \{1,\ldots,K\}$ with probability distribution $p(z_i)$ represents the classes. The (differential) entropy of this one variable is defined by
\begin{IEEEeqnarray}{lcl}
H(z_i) &=& - \E[\log p(z_i)]= - \sum_{k=1}^K  p(z_i = k) \log p(z_i = k) \IEEEnonumber \\
 &=& - \sum_{k=1}^K \pi_k \log \pi_k\cdot
\end{IEEEeqnarray} 
By assuming that the variables $\bz = (z_1,\ldots,z_n)$ are independent and identically distributed,
the whole entropy for $\bz$ in this i.i.d case is therefore additive and we have
\begin{IEEEeqnarray}{lcl}
H(\bz) &=& - \sum_{i=1}^n\sum_{k=1}^K \pi_k \log \pi_k\cdot
\label{eq: entropy of z}
\end{IEEEeqnarray}
The penalized log-likelihood function we propose to maximize is thus constructed by penalizing the observed data log-likelihood (\ref{eq: log-lik PRM}) by the entropy term (\ref{eq: entropy of z}), that is
\begin{equation}
\cJ(\lambda,\bsPsi) = \cL(\bsPsi) - \lambda H(\bz), \quad \lambda \geq 0
\end{equation}
which leads to the following penalized log-likelihood criterion: 
 {\small \begin{IEEEeqnarray}{lcl} 
\cJ(\lambda,\bsPsi) &=& \sum_{i=1}^n  \log \sum_{k=1}^K \pi_k \N (\by_{i};\bX_i \bsbeta_k,\sigma_k^2\Identity_m) +  \lambda \sum_{i=1}^n \sum_{k=1}^K  \pi_k \log \pi_k \nonumber \\
\label{eq: penalized log-lik for PRM}
\end{IEEEeqnarray}}where $\cL(\bsPsi)$ is the observed-data log-likelihood maximized by the standard EM algorithm for regression mixtures (see Equation (\ref{eq: log-lik PRM})) and $\lambda \geq 0$ is a parameter of control that controls the complexity of the fitted model.
This penalized log-likelihood function (\ref{eq: penalized log-lik for PRM}) we propose to optimize it in order to control the complexity of the model fit through roughness penalty $\lambda \sum_{i=1}^n \sum_{k=1}^K  \pi_k \log \pi_k$ in which the entropy $ -\sum_{i=1}^n \sum_{k=1}^K  \pi_k \log \pi_k$ measures the complexity of the fitted model for $K$ clusters. When
the entropy
 is large, the fitted model is rougher, and when it is small, the fitted model is smoother. The non-negative smoothing parameter $\lambda$ is for establishing a trade-off between closeness of fit to the data and a smooth fit. As $\lambda$ decreases, the fitted model tends to be less complex, and we get a
 smooth fit. However, when $\lambda$ increases, the result is a rough fit.
We discuss in the next section how to set this regularization  coefficient in an adaptive way.
%
 %
The next section shows how the penalized observed-data log-likelihood $\cJ(\lambda, \bsPsi)$ is maximized w.r.t the model parameters $\bsPsi$ by a robust EM algorithm for curve clustering.

\subsection{Robust EM algorithm for model-based curve clustering suing regression mixtures}
\label{ssec: Robust EM-MixReg}
Given an i.i.d training dataset of $n$ curves $\cD = ((\bx_1,\by_1),\ldots,(\bx_n, \by_n))$ 
the penalized log-likelihood (\ref{eq: penalized log-lik for PRM}) is iteratively maximized by using the following robust EM algorithm for model-based curve clustering.
Before giving the EM steps, we give the penalized complete-data log-likelihood, on which the EM formulation is relying, in this penalized case. The complete-data log-likelihood is given by
{\small \begin{IEEEeqnarray}{lll}
\cJ_c(\lambda, \bsPsi) & = & \sum_{i=1}^{n}\sum_{k=1}^{K} z_{ik} \log \big[\pi_k \N  (\by_{i};\bX_i\bsbeta_k ,\sigma^2_{k}\Identity_m)\big] \IEEEnonumber\\ 
& & + \lambda \sum_{i=1}^n \sum_{k=1}^K  \pi_k \log \pi_k \cdot 
 \label{eq: penalized complete log-lik for regression mixtures}
\end{IEEEeqnarray}}

After starting with an initial solution (see section \ref{ssec: initialization and stopping for Robust EM Mix-Reg} for the initialization strategy and stopping rule), the proposed algorithm  alternates between the two following steps until convergence. 
\subsubsection{E-step}
\label{ssec: M-step Robust EM-MixReg} 
This step computes the expectation of complete-data log-likelihood (\ref{eq: penalized complete log-lik for regression mixtures}) over the hidden variables $z_i$, given the observations $\cD$ and the current parameter estimation  $\bsPsi^{(q)}$, $q$ being the current iteration number: 
{\small
\begin{IEEEeqnarray}{lll}
& &\!\!\!\! Q(\lambda, \bsPsi;\bsPsi^{(q)}) =  \E\big[\cJ_c(\lambda, \bsPsi)|\cD;\bsPsi^{(q)}\big] \\
& = & \sum_{i=1}^{n}\sum_{k=1}^{K}\tau_{ik}^{(q)} \log \big[\pi_k \N  (\by_i;\bX_i \bsbeta_{k},\sigma^2_{k}\Identity_m)\big]+ \lambda \sum_{i=1}^n \sum_{k=1}^K  \pi_k \log \pi_k \IEEEnonumber 
\label{eq: Q-function for the regression mixtures}
\end{IEEEeqnarray}}which simply consists in computing the posterior cluster probabilities 
$\tau_{ik}^{(q)}$  given by: 
\begin{equation} 
\tau_{ik}^{(q)} =  \frac{\pi_k^{(q)} \N\big(\by_i;\bX_i \bsbeta^{(q)}_{k},\sigma^{2(q)}_{k}\Identity_m\big)}{\sum_{h=1}^K \pi_{h}^{(q)} \N(\by_i;\bX_i \bsbeta^{(q)}_{h},\sigma^{2(q)}_{h}\Identity_m)}\cdot
\label{eq: Robust-MixReg post prob PRM}
\end{equation}

\subsubsection{M-step}
\label{ssec: M-step Robust EM-MixReg} 
This step updates the value of the parameter vector $\bsPsi$ by maximizing the 
the $Q$-function (\ref{eq: Q-function for the regression mixtures}) with respect to $\bsPsi$, that is: 
$\bsPsi^{(q+1)} = \arg \max_{\bsPsi} Q(\lambda, \bsPsi;\bsPsi^{(q)}).$  
It can be shown that this maximization can be performed by separate  maximizations w.r.t the mixing proportions $(\pi_{1},\ldots,\pi_{K})$ subject to the constraint $\sum_{k=1}^{K} \pi_{k} = 1$, 
and w.r.t the regression parameters $\{\bsbeta_{k},\sigma^2_{k}\}$.
The mixing proportions updates are obtained by maximizing the function
\begin{equation}
Q_{\pi}(\lambda; \bsPsi^{(q)}) = \sum_{i=1}^{n}\sum_{k=1}^{K}\tau_{ik}^{(q)} \log \pi_k + \lambda \sum_{i=1}^n \sum_{k=1}^K  \pi_k \log \pi_k \nonumber
\label{eq: J(pik) pinalized log-lik PRM} 
\end{equation}w.r.t the mixing proportions $(\pi_{1},\ldots,\pi_{K})$ subject to the constraint $\sum_{k=1}^{K} \pi_{k} = 1$. This can be solved using Lagrange multipliers (see Appendix or \cite{Robust_EM_GMM}) and the obtained updating formula is  given by:
\begin{equation}
\pi_{k}^{(q+1)} = \frac{1}{n}\sum_{i=1}^n \tau_{ik}^{(q)} + \lambda \pi_{k}^{(q)}\left(\log \pi_{k}^{(q)} - \sum_{h=1}^K\pi_{h}^{(q)}\log \pi_{h}^{(q)}\right)\cdot
\label{eq: Robust EM-MixReg pi_k update}
\end{equation}
%
We notice here that in the update of the mixing proportions (\ref{eq: Robust EM-MixReg pi_k update}) the update is close to the standard EM update $\big(\frac{1}{n}\sum_{i=1}^n \tau_{ik}^{(q)}$ see Eq. (\ref{eq: EM-MixReg pik update})$\big)$ for very small value of $\lambda$. However, for a large value of $\lambda$, the penalization term will play its role in order to make clusters competitive and thus allows for discarding illegitimate clusters and enhancing actual clusters.
Indeed, in the updating formula (\ref{eq: Robust EM-MixReg pi_k update}), we can see that 
for cluster $k$
if
\begin{equation}
\left(\log \pi_{k}^{(q)} - \sum_{h=1}^K\pi_{h}^{(q)}\log \pi_{h}^{(q)}\right) > 0
\label{eq: step of pik update Robust EM-MixReg}
\end{equation}
that is, for the (logarithm of the) current proportion $\log \pi_{k}^{(q)}$, the entropy of the hidden variables is decreasing, and therefore the model complexity tends to be stable, 
the cluster $k$ has therefore to be enhanced. This explicitly results in the fact that the update of the $k$th mixing proportion $\pi_{k}^{(q+1)}$ in (\ref{eq: Robust EM-MixReg pi_k update}) will increase. 
%
On the other hand, if (\ref{eq: step of pik update Robust EM-MixReg}) is less than $0$, the cluster proportion will therefore decrease as is not very informative in the sense of the entropy.

Finally, the penalization coefficient $\lambda$ has to be set as described previously in such a way to be large for enhancing competition when the proportions are not increasing enough 
from one iteration to another. In this case, the robust algorithm plays its role for estimating the number of clusters (which is decreasing in this case by discarding small illegitimate clusters). We note that here a cluster $k$ can be discarded if its proportion is less than $\frac{1}{n}$, that is $\pi_{k}^{(q)}<\frac{1}{n}$.
On the other hand,  $\lambda$ has to become small when the proportions are sufficiently increasing as the learning proceeds and the partition can therefore be considered as stable. In this case, the robust EM algorithm tends to have the same behavior as the stand EM described in section \ref{ssec: EM_MixReg}.
The regularization coefficient First $\lambda$ is set as follows (similarly as described in \cite{Robust_EM_GMM} for Gaussian mixtures). First, it can be set in $[0,1]$ to prevent very large values. Furthermore, the following adaptive formula for $\lambda$, as shown in \cite{Robust_EM_GMM} for the case of Gaussian mixtures for multivariate data clustering, can be used to adapt it as the learning proceeds:  
\!\! {\small \begin{IEEEeqnarray}{lll}
&& \lambda^{(q+1)}= \min \IEEEnonumber\\
&& \left\{\frac{\sum_{k=1}^K \exp\left(\eta n |\pi_{k}^{(q+1)} - \!  \pi_{k}^{(q)}|\right)}{K}, \frac{1- \max_{k=1}^K\left(\frac{\sum_{i=1}^n\tau^{(q)}_{ik}}{n}\right)}{-\pi_{k}^{(q)} \sum_{k=1}^K \pi_{k}^{(q)} \log \pi_{k}^{(q)}}\right\} \IEEEnonumber\\
\label{eq: lambda update Robust EM-MixReg}
\end{IEEEeqnarray}}%
where $\eta$ can be set as $\min(1, 0.5^{\lfloor\frac{m}{2}-1\rfloor})$, $m$ being the number of observations per curve and $\lfloor x \rfloor$ denotes the largest integer that is no more than $x$.
%
%

The maximization w.r.t the regression parameters however consists in separately maximizing for each class $k$ the function
{\small \begin{IEEEeqnarray}{lll}
&&\!\!\!\!\!\!\!\!  Q_{\bsPsi_k}(\lambda,\bsbeta_k,\sigma^2_k; \bsPsi^{(q)}) = \sum_{i=1}^{n} \tau_{ik}^{(q)} \log  \N  (\by_i;\bX_i \bsbeta_{k},\sigma^2_{k}\Identity_m) \IEEEnonumber \\
&=&\sum_{i=1}^{n}\tau^{(q)}_{ik}\left[-\frac{m}{2}\log 2 \pi - \frac{m}{2}\log \sigma^2_{k} - \frac{1}{2 \sigma^2_{k}} \parallel \by_i - \bX_i\bsbeta_k\parallel^2\right] \nonumber 
\end{IEEEeqnarray}}w.r.t $(\bsbeta_{k},\sigma^2_{k})$.
This maximization w.r.t the regression parameters consists in performing analytic solutions of  weighted least-squares problems where the weights are the posterior cluster probabilities $\tau_{ik}^{(q)}$. The updating formula are given by:
\begin{IEEEeqnarray}{lll}
\bsbeta_k^{(q+1)}  &=& \Big[\sum_{i=1}^{n}\tau^{(q)}_{ik} \bX^T_i\bX_i \Big]^{-1} \sum_{i=1}^{n}\tau^{(q)}_{ik} \bX_i^T \by_i
\label{eq: Robust EM-MixReg beta_k update}
\end{IEEEeqnarray} 
\vspace{-.4cm}
\begin{IEEEeqnarray}{lll}
\sigma_k^{2(q+1)} &=& \frac{1}{m\sum_{i=1}^{n}\tau^{(q)}_{ik}} \sum_{i=1}^{n}\tau^{(q)}_{ik} \parallel \by_i - \bX_i\bsbeta_k\parallel^2 
\label{eq: Robust EM-MixReg sigma_k update}
\end{IEEEeqnarray}where the posterior cluster probabilities $\tau^{(q)}_{ik}$ given by (\ref{eq: Robust-MixReg post prob PRM})
are computed using the mixing proportions derived in (\ref{eq: Robust EM-MixReg pi_k update}).
Then, once the model parameters have been estimated, a partition of the data into $K$ clusters is then computed by maximizing the posterior cluster probabilities (\ref{eq: Robust-MixReg post prob PRM}).
%
\subsection{Initialization strategy and stopping rule}
\label{ssec: initialization and stopping for Robust EM Mix-Reg} 
The initial number of clusters is $K^{(0)}= n$, $n$ being the total number of curves and the initial mixing proportions are $\pi_k^{(0)}= \frac{1}{K^{(0)}}$, ($k=1,\ldots, K^{(0)}$). Then, to initialize the regression parameters $\bsbeta_k$ and the noise variances $\sigma_k^{2(0)}$ ($k=1,\ldots, K^{(0)}$), we fitted a polynomial regression models on each curve $k$, ($k=1,\ldots, K^{(0)}$) and the initial values of the regression parameters are therefore given by
$\bsbeta_k^{(0)}  = \Big(\bX^T\bX_k \Big)^{-1}\bX_k \by_k$
and the noise variance can be deduced as
$\sigma_k^{2(0)} = \frac{1}{m} \parallel\by_k - \bX_k\bsbeta_k^{(0)}\parallel^2$. However, to avoid singularities at the starting point, we set $\sigma_k^{2(0)}$ as a middle value in the following sorted range $\parallel\by_i - \bX\bsbeta_k^{(0)}\parallel^2$ for $i=1,\ldots,n$.
 
The proposed EM algorithm is stopped when the maximum variation of the estimated regression parameters between two iterations $\max_{1\leq k \leq K^{(q)}} \parallel \bsbeta_k^{(q+1)} - \bsbeta_k^{(q)}\parallel$ was less than a threshold $\epsilon$ (e.g., $10^{-6}$). 
The pseudo code \ref{algo: proposed Robust EM-MixReg} summarizes the proposed robust EM algorithm for model-based curve clustering using regression mixtures. 
\begin{algorithm}[htbp]
\caption{\label{algo: proposed Robust EM-MixReg} Pseudo code of the proposed Robust EM algorithm for regression mixtures.}
{\bf Inputs:} Set of curves $\cD = ((\bx_1,\by_1),\ldots,(\bx_n,\by_n))$ and polynomial degree $p$
\begin{algorithmic}[1]
\STATE $\epsilon \leftarrow 10^{-6}$; $q \leftarrow 0$;  $converge \leftarrow 0$ 

\verb|//Initialization:| 
\STATE $\bsPsi^{(0)}= (\pi_1^{(0)},\ldots,\pi_K^{(0)}, \bsPsi_1^{(0)},\ldots,\bsPsi_K^{(0)})$; $K^{(0)} = n$

\FOR{$k=1,\ldots,K^{(q)}$}
 \STATE Compute $\tau_{ik}^{(q)}$ for $i=1,\ldots,n$  using Equation (\ref{eq: Robust-MixReg post prob PRM})  
 \STATE Compute $\pi_k^{(q)}$ using Equation (\ref{eq: Robust EM-MixReg pi_k update})
\ENDFOR

\verb|//EM loop:| 
\WHILE{{\small $(! \ converge)$}}
		\FOR{$k=1,\ldots,K^{(q)}$}
		\STATE Compute $\tau_{ik}^{(q)}$ for $i=1,\ldots,n$  using Equation (\ref{eq: Robust-MixReg post prob PRM})  
	\ENDFOR
	\FOR{$k=1,\ldots,K^{(q)}$} 
		\STATE Compute $\pi_k^{(q+1)}$ using Equation (\ref{eq: Robust EM-MixReg pi_k update})
	\ENDFOR		
	 \STATE Compute $\lambda^{(q+1)}$ using Equation (\ref{eq: lambda update Robust EM-MixReg})
	  ; Discard illegitimate clusters with small proportions  $\pi_k^{(q)}< \frac{1}{n}$
	  ; Set $K^{(q+1)} = K^{(q)} - \#\{\pi_k^{(q)}|\pi_k^{(q)}< \frac{1}{n}\}$
	  ; Adjust $\tau_{ik}^{(q+1)}$ and $\pi_k^{(q+1)}$ to make their columns sum to one 
	\FOR{$k=1,\ldots,K^{(q)}$} 
		\STATE Compute $\bsbeta_k^{(q+1)}$ using Equation (\ref{eq: Robust EM-MixReg beta_k update})
		\STATE Compute $\sigma_k^{2(q+1)}$ using Equation (\ref{eq: Robust EM-MixReg sigma_k update})
	\ENDFOR
	\IF{$\max_{1\leq k \leq K^{(q)}} \parallel \bsbeta_k^{(q+1)} - \bsbeta_k^{(q)}\parallel <\epsilon$}
	 \STATE $converge =1$
	\ENDIF 
	\STATE $q \leftarrow q+1$ 
\ENDWHILE
\end{algorithmic}
{\bf Outputs:} $K= K^{(q)}, \quad \hat{\bsPsi} = \bsPsi^{(q)}$, \quad $\hat{\tau}_{ik}=  \tau_{ik}^{(q)}$ 
\end{algorithm}

\section{Simulation study}  
\label{sec: Experiments} 
\begin{figure}[!ht]
 \centering  
 \includegraphics[width=4cm, height=2.9cm]{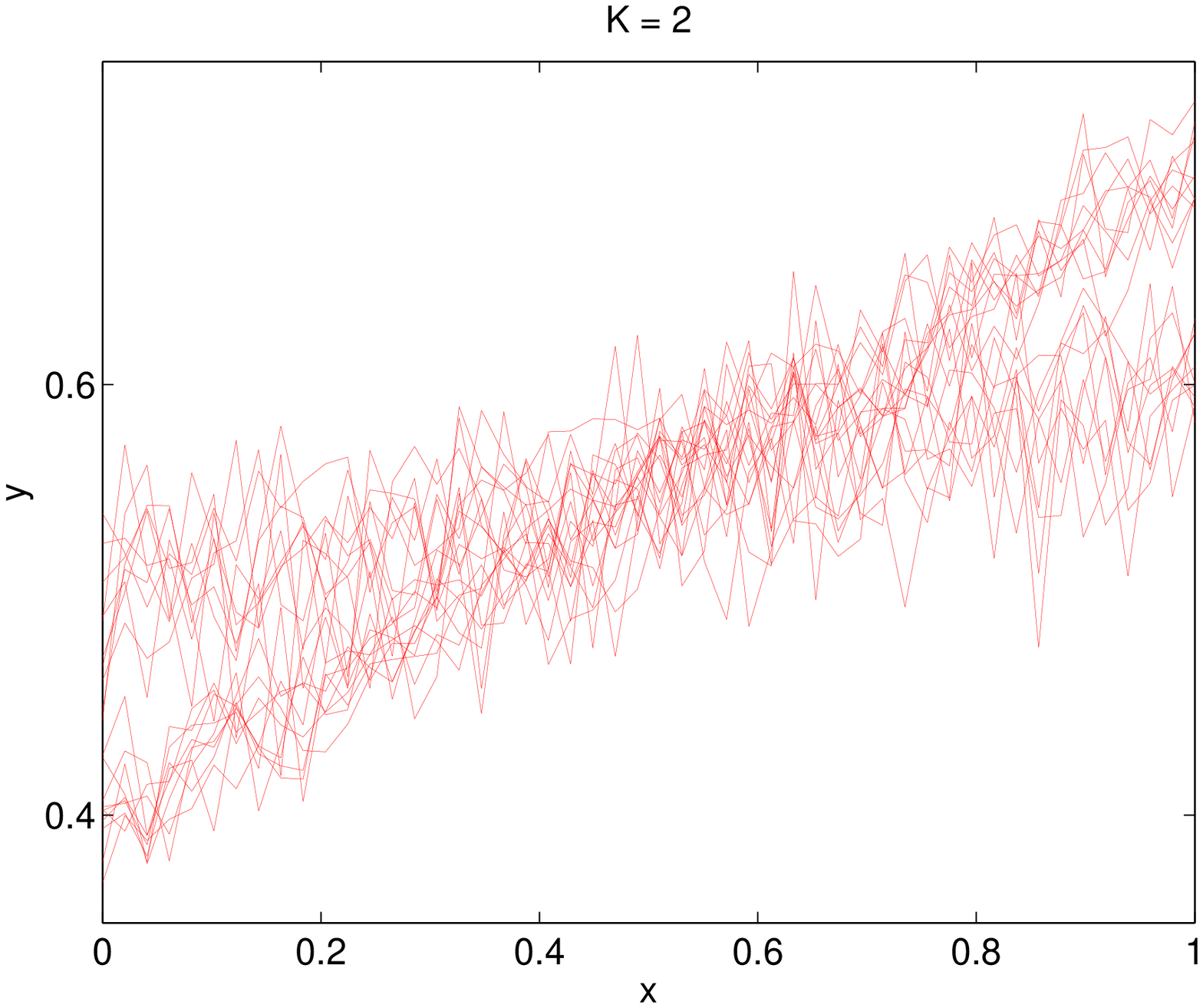} 
 \includegraphics[width=4cm, height=2.9cm]{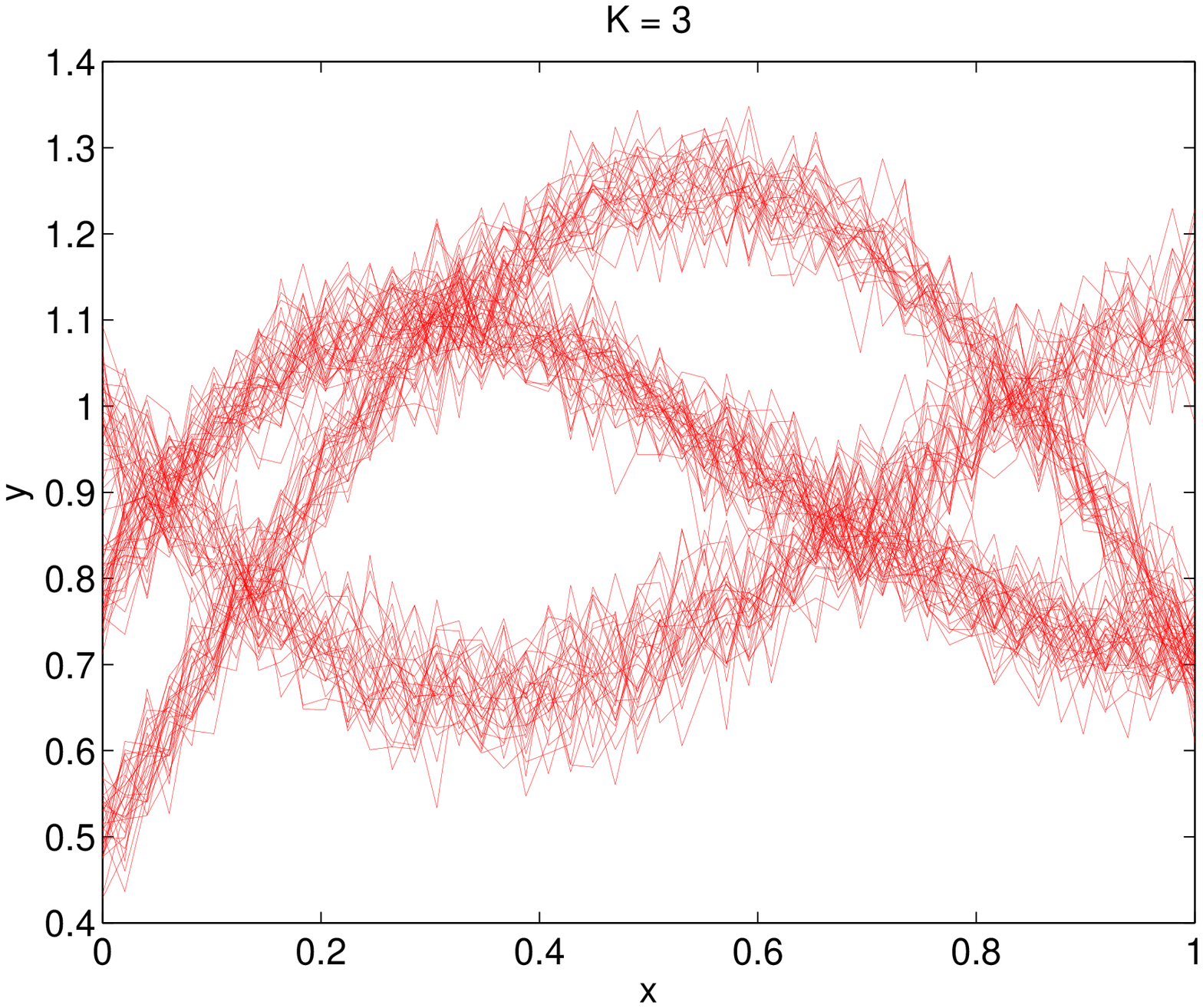}  
 \caption{\label{fig: simulated data (2 situations)}Simulates curves for the two situations}
\end{figure}

This section is dedicated to the evaluation of the proposed approach on simulated curves.
 We consider linear and  non-linear arbitrary curves (not simulated according to the model) as presented respectively in Fig. 
   \ref{fig: simulated data (2 situations)}.
The first situation consists in a two-class problem. 
The simulated dataset consists of $n=20$ curves, each curve consists of $m=50$ observationas generated as a linear function corrupted by a Gaussian noise as follows. For the $i$th curve ($i=1,\ldots,n$), the $j$th observation ($j=1,\ldots,m$) is generated as follows:
\begin{itemize}
\item $y_{ij}= 0.3\, x_{ij} + 0.4 + \sigma_1\,\epsilon_{ij}$;
\item $y_{ij}= 0.1\, x_{ij} + 0.5 + \sigma_2\,\epsilon_{ij}$.
\end{itemize}respectively for class 1 and class 2, where $x$ are linearly equally spaced points in  the range $[0,1]$,
with $\sigma_1 = 0.02$ and $\sigma_2 = 0.03$ are the corresponding noise standard deviations, and $\epsilon_{ij} \sim \N(0,1)$ are standard zero-mean unit variance Gaussian variables.
The two classes have the same proportion. 
Figure \ref{fig: simulated data (2 situations)} (top) shows this dataset where the true value of the number of clusters is $K=2$.

\begin{figure*}[ht]
 \centering  
 \includegraphics[width=4.6cm, height=3.cm]{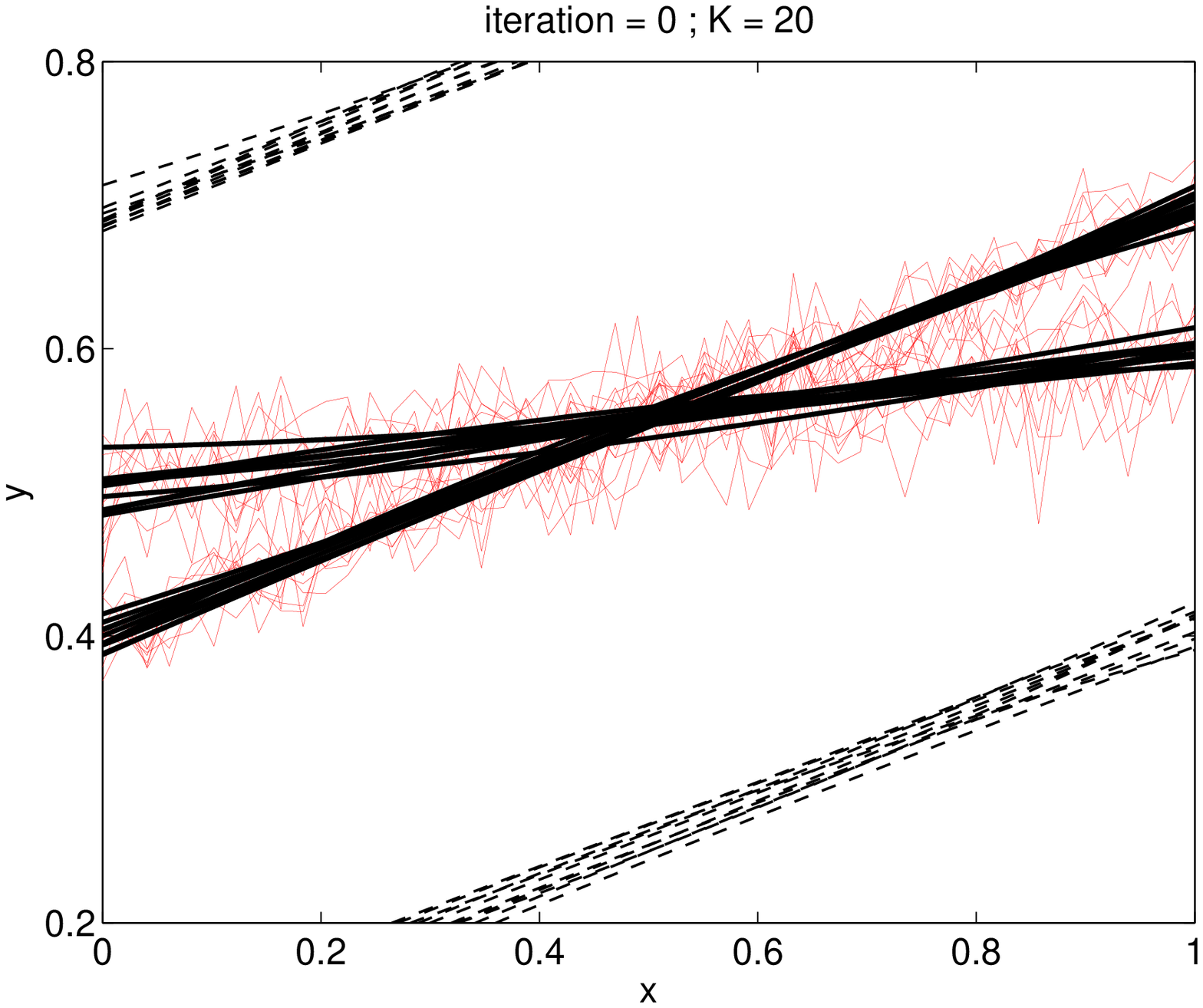} 
 \includegraphics[width=4.6cm, height=3.1cm]{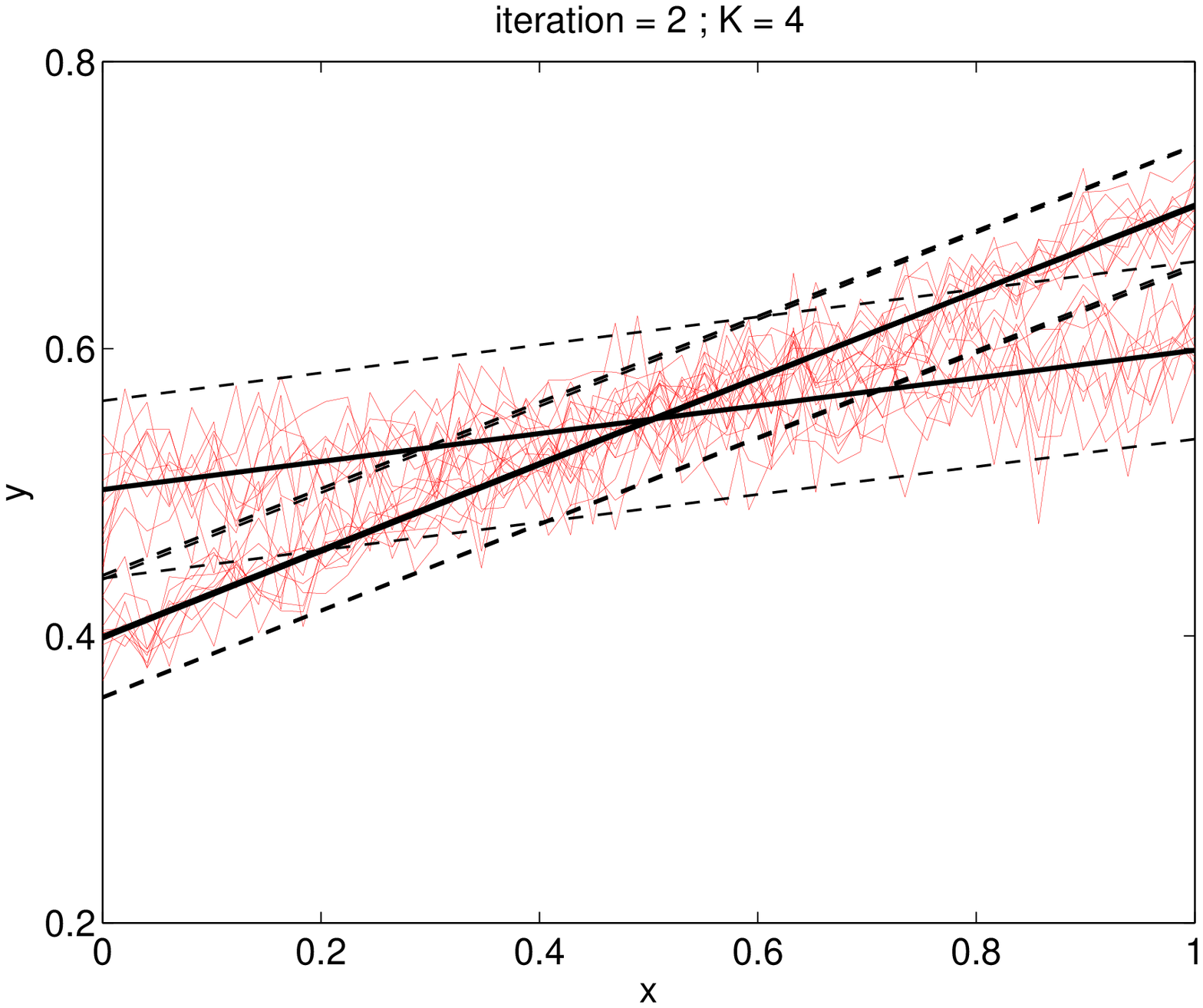} 
 \includegraphics[width=4.6cm, height=3.1cm]{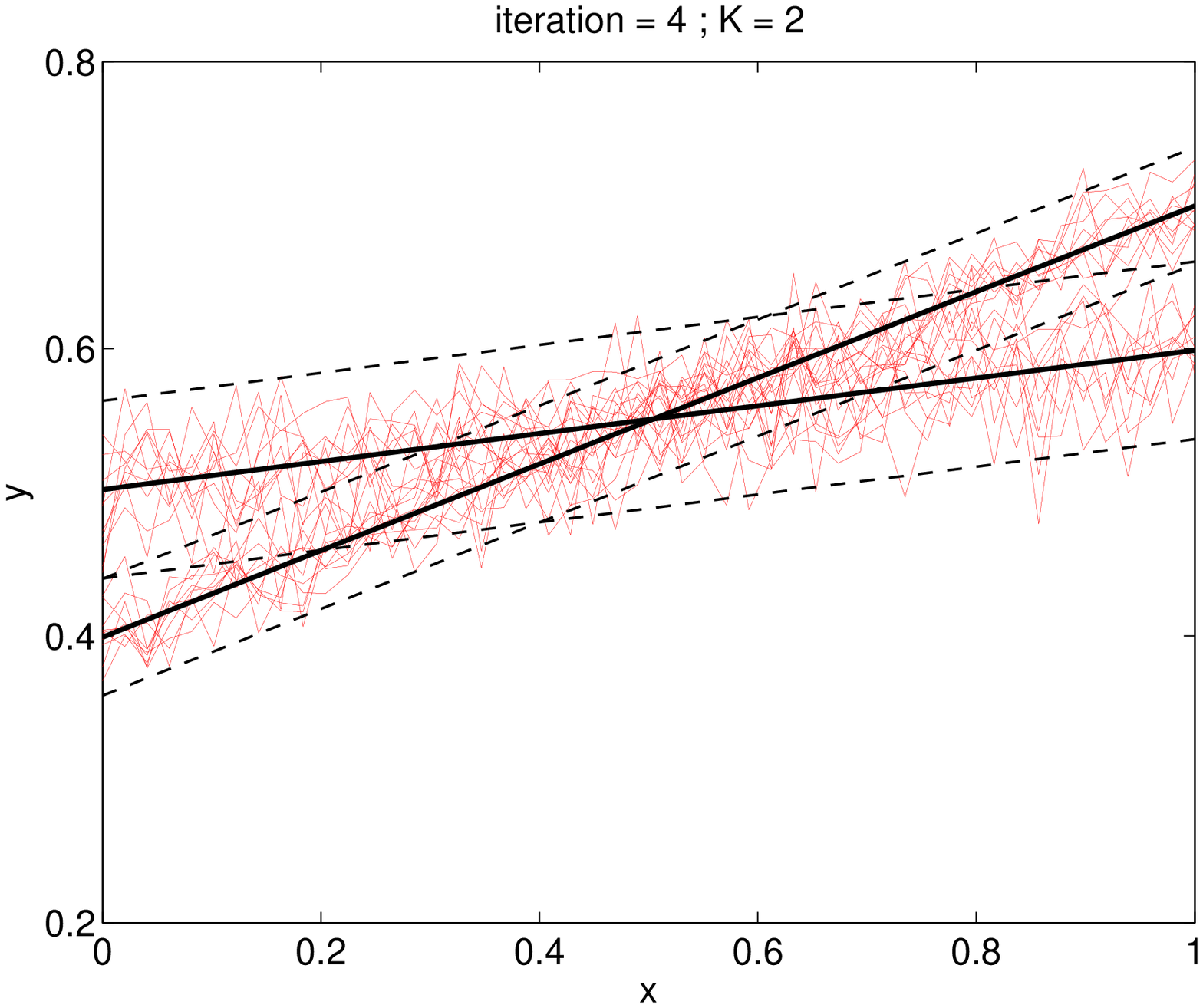}
 \caption{\label{fig: EM-MixReg results linear curves}The clustering results obtained with the proposed robust EM-MixReg algorithm for the first set of curves shown in Fig. \ref{fig: simulated data (2 situations)}.} 
\end{figure*}

\begin{figure*}[htbp]
 \centering 
 \includegraphics[width=4.6cm, height=3.1cm]{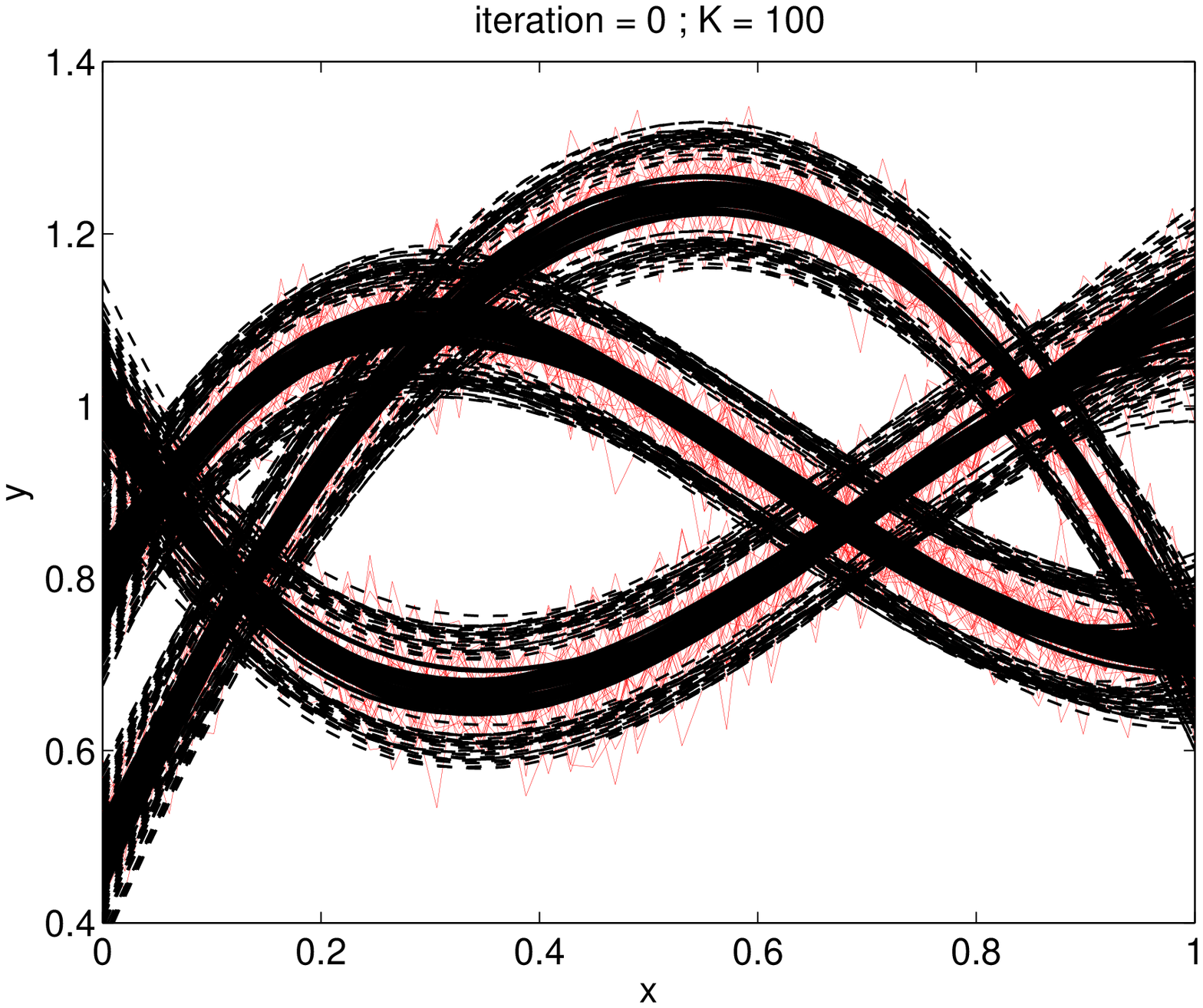} 
 \includegraphics[width=4.6cm, height=3.1cm]{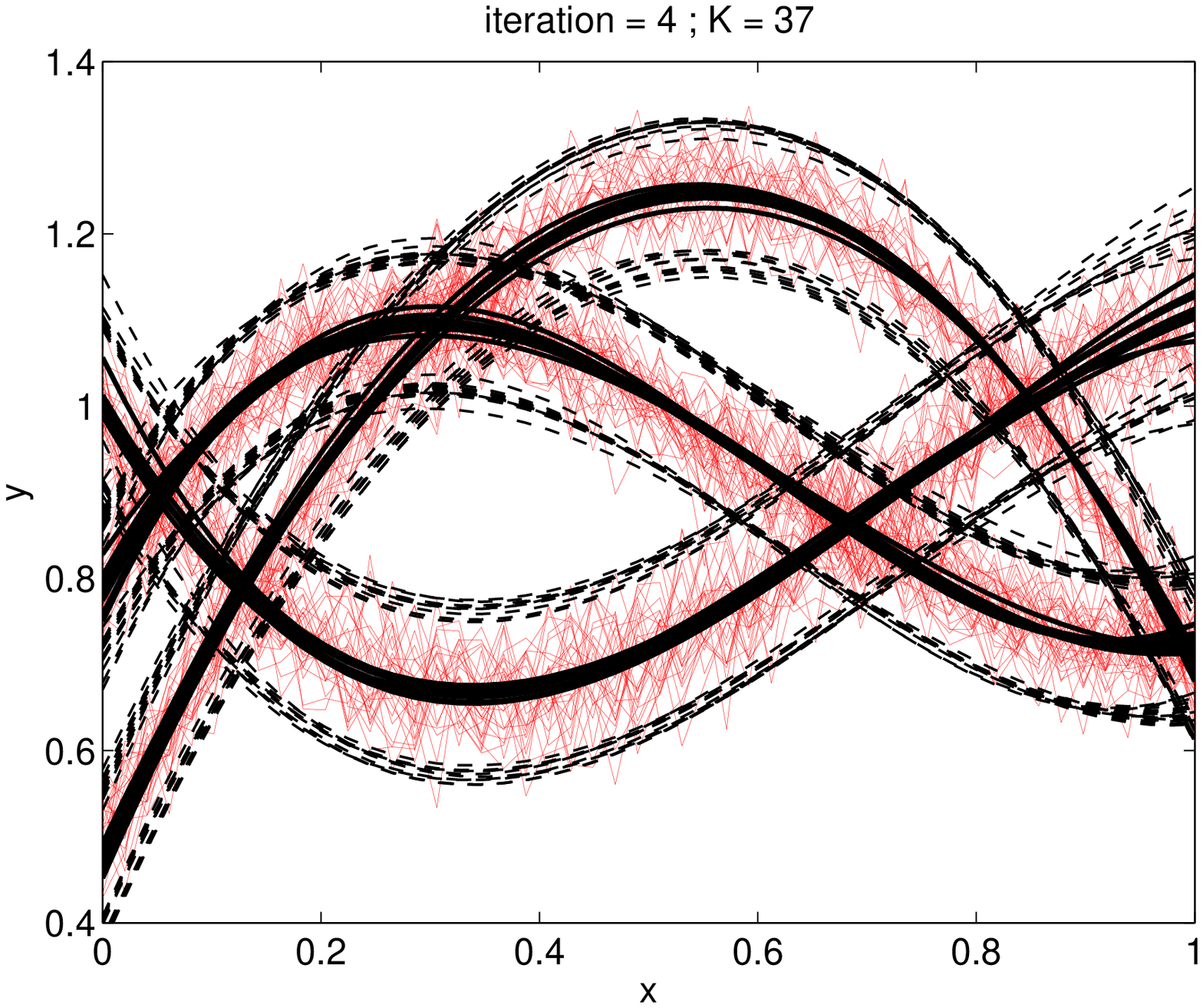}
 \includegraphics[width=4.6cm, height=3.1cm]{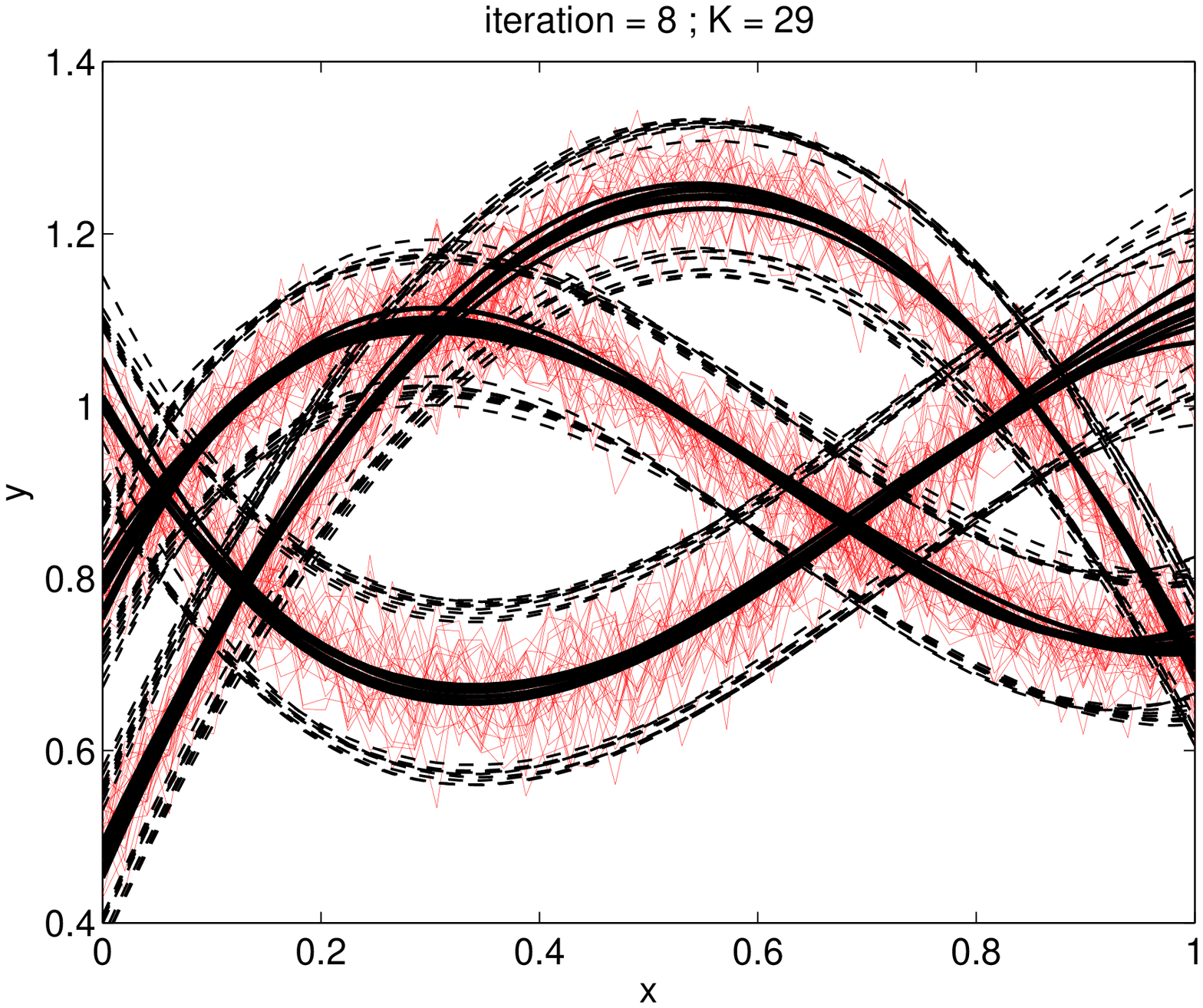}
\\
 \includegraphics[width=4.6cm, height=3.1cm]{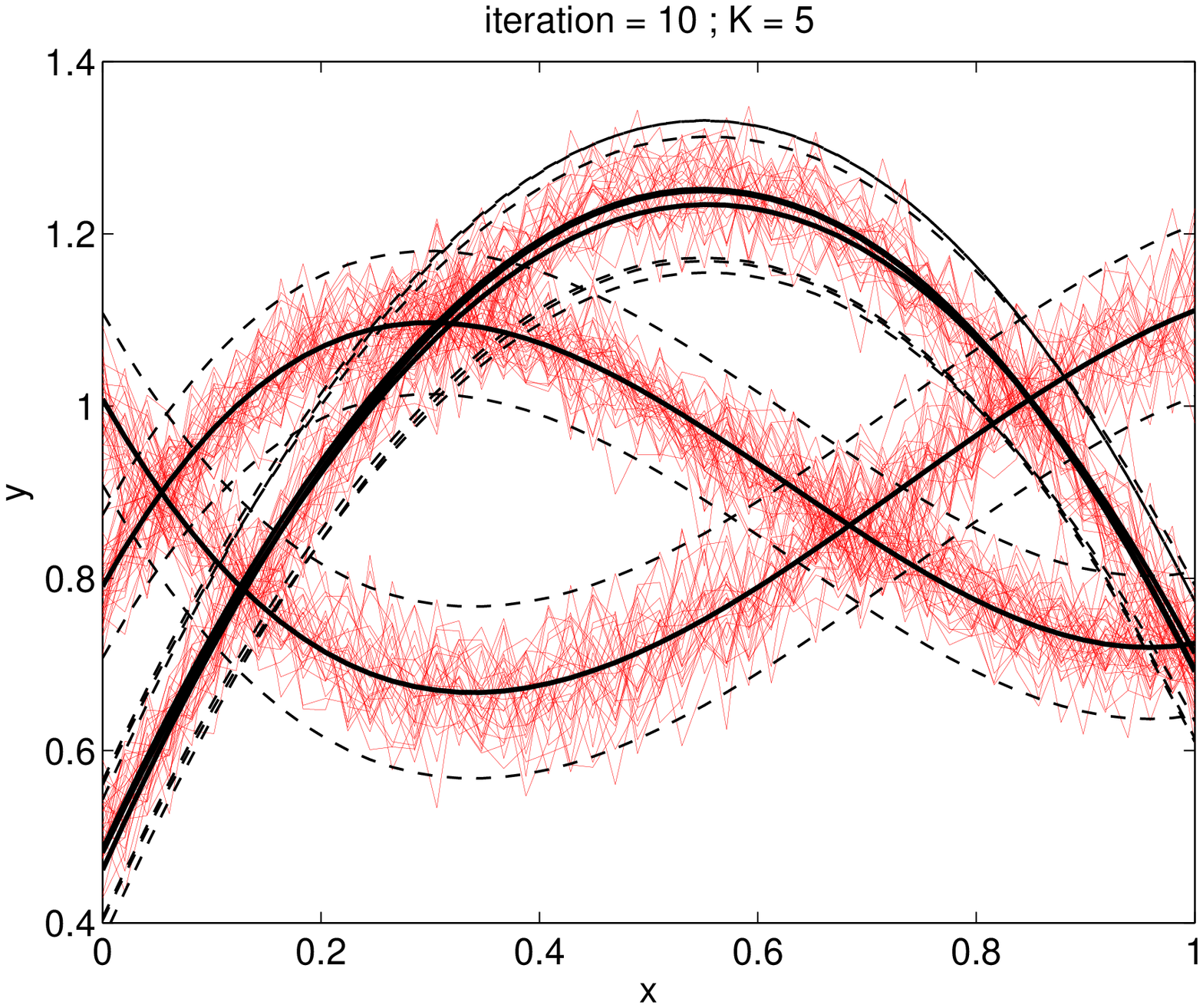}
 \includegraphics[width=4.6cm, height=3.1cm]{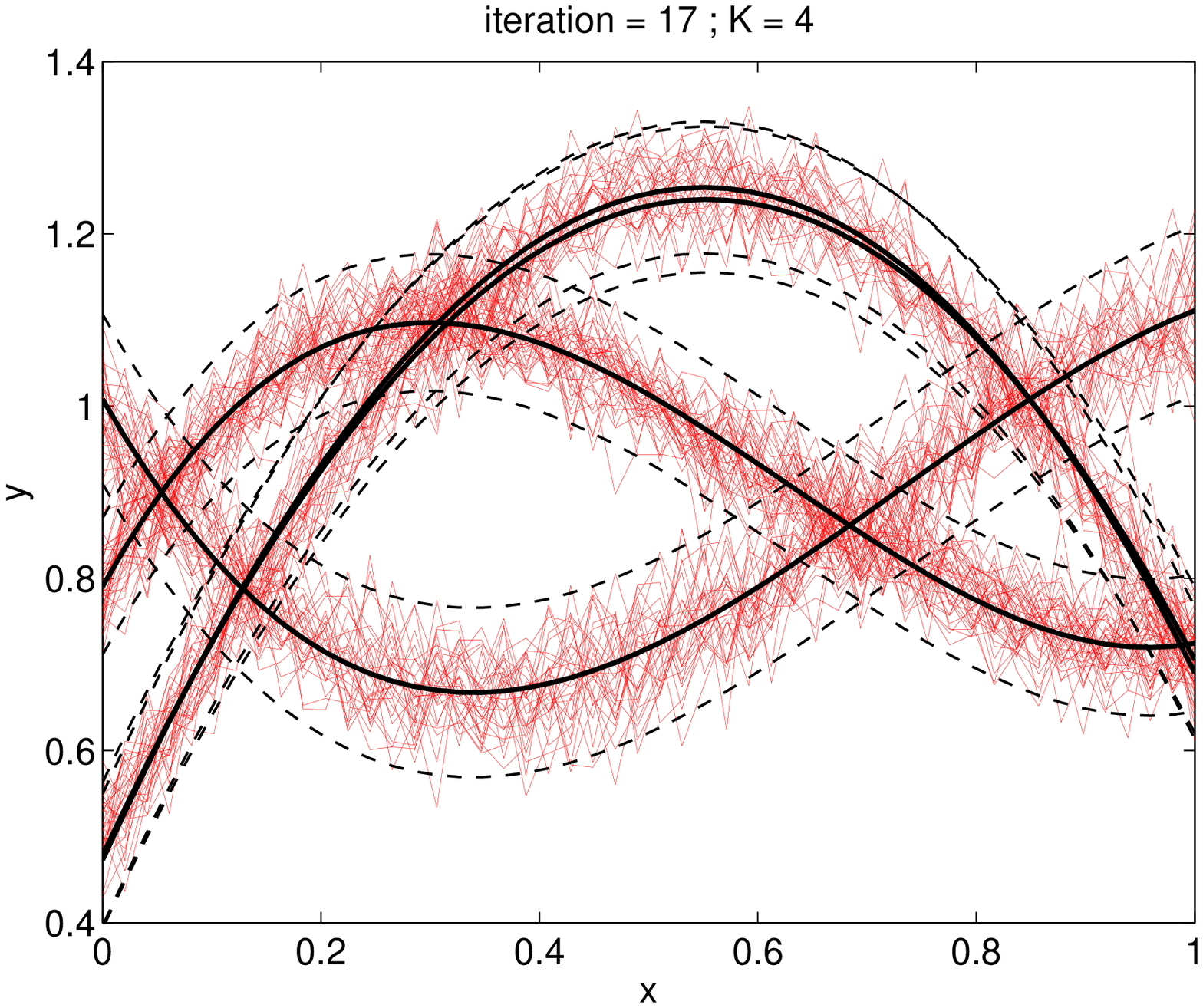}
 \includegraphics[width=4.6cm, height=3.1cm]{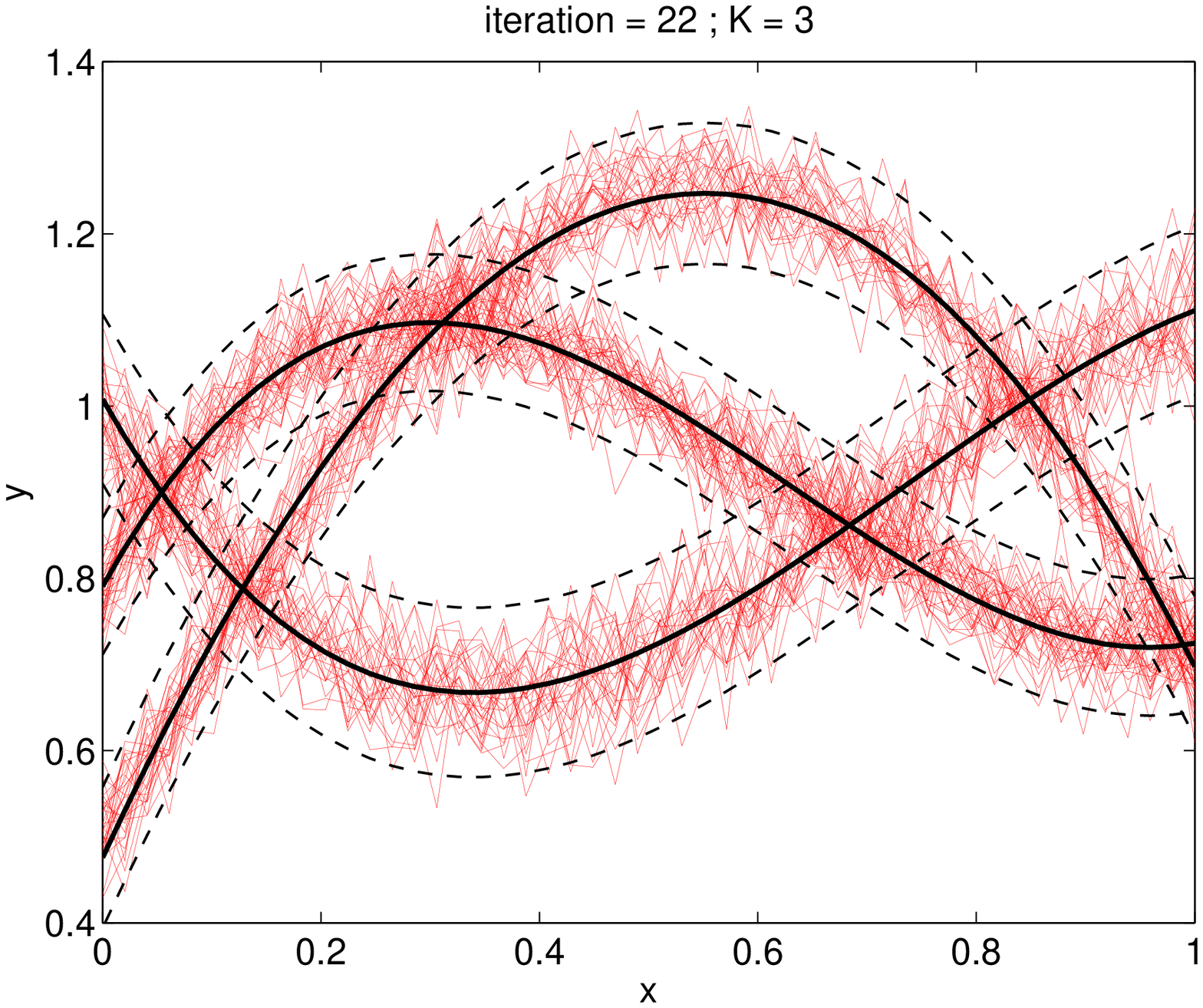}  
 \caption{\label{fig: EM-MixReg results non-linear curves}The clustering results obtained with the proposed robust EM-MixReg algorithm for the second set of curves shown in Fig. \ref{fig: simulated data (2 situations)}.} 
\end{figure*}

The second situation represents a three-class problem. 
The simulated dataset consists of $n=100$ arbitrary non-linear curves, each curve consists of $m=50$ observations generated as follows. For the $i$th curve ($i=1,\ldots,n$), the $j$th observation ($j=1,\ldots,m$) is generated as follows:
\begin{itemize}
\item $y_{ij}=  0.8 + 0.5\,\exp(-1.5\, x)\,\sin(1.3 \pi\, x) + \sigma_1\,\epsilon_{ij}$;
\item $y_{ij}=  0.5 + 0.8\,\exp(-x)\,\sin(0.9 \pi\, x) + \sigma_2\,\epsilon_{ij}$;
\item $y_{ij}=  1 + 0.5\,\exp(-x)\,\sin(1.2 \pi\, x) + \sigma_3\,\epsilon_{ij}$. 
\end{itemize} with $\sigma_1 = 0.04$, $\sigma_2 = 0.04$ and $\sigma_3 = 0.05$.
The classes have respectively proportions $\pi_1 = 0.4, \pi_2 = 0.3, \pi_3 = 0.3$.
%
%
Figure \ref{fig: EM-MixReg results linear curves} shows the obtained results for the first set of linear curves obtained with a linear regression mixture. 
For the first dataset, after only two iterations, the majority of illegitimate clusters are discarded and the algorithm converges in 4 iterations and provides the correct clustering results with the actual number of clusters.
%
%
%
%
%
Figure \ref{fig: EM-MixReg results non-linear curves} shows the obtained resulted for the second set of arbitrary non-linear curves obtained with a polynomial regression mixture ($p=3$). 
The figures show that the algorithm started with a number of clusters equal to the number of curves. 
It can be seen that for the second data set of non-linear curves, the proposed algorithm also provides accurate results. After starting with a number of clusters $K=100$, the number of clusters decreases rapidly form 100 to 27 after only four iterations. Then the algorithm converges after 22 iterations and provides the actual number of clusters with precise clustering results.
  

\section{Conclusion}
\label{sec: conclusion}
In this paper, we presented a new EM algorithm for model-based curve clustering. It optimizes a penalized observed-data log-likelihood using the entropy of the hidden structure.
The proposed algorithm overcome both the problem of sensitivity to initialization and determining the optimal number of clusters for standard EM for regression mixtures.
The experimental results on simulated data demonstrates the potential benefit of the proposed approach for curve clustering.
Future work will concern additional experiments on real data including temporal curves. 

\bibliographystyle{IEEEtran}
\bibliography{references} 
 
\end{document}

%% file: mymathsym.tex
\newcommand{\bO}{\mathbf{0}}

\newcommand{\bx}{\mathbf{x}}
\newcommand{\bX}{\mathbf{X}}
\newcommand{\by}{\mathbf{y}}

\newcommand{\bz}{\mathbf{z}}

\newcommand{\bsx}{\boldsymbol{x}}

\newcommand{\cL}{\mathcal{L}}

\newcommand{\cD}{\mathcal{D}}

\newcommand{\cJ}{\mathcal{J}}

\newcommand{\bsmu}{\boldsymbol{\mu}}

\newcommand{\bsbeta}{\boldsymbol{\beta}}

\newcommand{\bsSigma}{\boldsymbol{\Sigma}}

\newcommand{\bsepsilon}{\boldsymbol{\epsilon}}

\newcommand{\bsPsi}{\boldsymbol{\Psi}}

\newcommand{\Identity}{\textbf{I}}

\newcommand{\E}{\mathbb{E}}

\newcommand{\R}{\mathbb{R}}
	
\newcommand{\N}{\mathcal{N}}